\documentclass[twocolumn,showpacs,preprintnumbers,amsmath,amssymb]{revtex4}
\usepackage{graphicx}
\usepackage{dcolumn}
\usepackage{bm}

\begin{document}

\preprint{APS/123-QED}

\title{  The boundary of G\"odel's spacetime and the chronology protection conjecture }
\author{P.Pitanga}
\affiliation{Universidade Federal do Rio de Janeiro. Instituto de
F\'{\i}sica. Caixa Postal 68528 \\
Cidade Universit\'aria. 21945-970. Rio de Janeiro, Brazil }
\email{pitanga@if.ufrj.br}
\date{\today}
\begin{abstract}
 We present a homogenous anisotropic conformal spacetime manifold that provide an example of Hawking's chronology protection conjecture in three-dimensional gravity theory. The solution is based upon the fact that the seven-dimensional group of the automorphism of the Heisenberg motion group $H^1 \times U(1)$, modulo discrete sub-group ${\Gamma}$, is the symmetry group of the sub-Riemannian (SR)-manifold, boundary  of the Cauchy-Riemann  (CR)-manifold, allowing the existence of positive mass, momentum, angular-momentum and timelike-translation. It is shown that many mirror symmetric self-similar G\"odel's surfaces are hidden behind a Cauchy spacelike  surface so that causality violation is not visible from outside. 
\end{abstract}
\maketitle 
 In 1949, G\"odel \cite {KU} discovered a solution for Einstein's field equations in which causality is violated by the exis\-tence of  periodic world lines that run back into themselves. The periods are given by  integrals over the proper time differential $ds$. These cycles are known as closed timelike curves (CTCs). According to Tipler \cite {TI} a massive infinite  rotating cylinder should create a frame dragging effect in the spacetime giving rise to CTCs. Tipler suggested, without proof, that in a finite rotating cylinder CTCs would arise allowing travel into the past. CTCs also arise in  Misner spaces \cite {MI},wormholes \cite {TOR} and in  Gott's  two-strings spacetime \cite {GOT,CURT}. 

The non existence of CTCs in acceptable spacetimes, was demonstrated by Deser, Jackiw and 't Hooft \cite {DD} in 1992. In the same year, Hawking \cite {HH}  showed that, according to the ge\-ne\-ral theory of relativity, it is impossible to build a time machine in finite regions of the spacetime without curvature singularities, due to the presence of negative energy states. Hawking posited the existence of a Cauchy-Riemann spacelike surface, with a compact horizon, so that causality violation could never be observed from  outside. This conjecture is known as {\it Chronology Protection Conjecture}. For G\"odel's universe and CPC, see \cite{HW,OS,BO,MIG,HA,SC,SQ,CU}

The aim of this letter is to  provide the appropriate geometrical setting to give an example of CPC, by using G\"odel's solution. For our purpose it is sufficient to a\-na\-ly\-ze the $(2+1)$-dimensional G\"odel spacetime manifold with constant angular velocity ${\omega}_0$, and negative cosmological constant ${\Lambda}$. By spacetime manifold, $({\cal M},d)$, we mean the space of points ${\cal M}$, with metric distance $d$. Two models $({\cal M}_1,d_1)$ and $({\cal M}_2,d_2)$ are equivalents if there exist a {\it group of diffeomorphism } $f :({\cal M}_1,d_1)\rightarrow ({\cal M}_2,d_2)$, such that $d_2(f(x),f(y))=d_1(x,y)$. In this case we shall work with just one representative, as in \cite {HW}.
 
First of all we show that the Cauchy spacelike surface of G\"odel's spacetime is {\it equivalent} to the compact homogeneous manifold of the reduced Heisenberg motion group ${\overline G}=G/U(1)=H^1/{\Gamma}$. Thus, the sub-Riemannian Cauchy  problem is solved in the cotangent  space, $T^{*}{\cal M}$, of ${\overline G}$. 

{\it A Sub-Riemannian manifold} is  a smooth ma\-ni\-fold, ${\cal M}$, equipped with a smoothly varying positive definite quadratic
form  on a sub-bundle ${\cal S}$ of the tangent space $T{\cal M}$, where ${\cal S}$ is assumed to be {\it bracket generating} (sections of ${\cal S}$ together with all brackets generate $T{\cal M}$ as a module over the functions on ${\cal M}$). In local coordinates we write $g^{{\mu}{\nu}}(x)$ for the sub-Riemannian metric. However there is no analogue of {\it affine connection}, ${\Gamma}^{\lambda}_{{\mu}{\nu}}(x)$, since $g^{{\mu}{\nu}}(x)$ {\it is never invertible}. All admissible curves  must be {\it horizontal}, that is tangent to ${\cal S}$. The SR-geodesics, in general, are not shortest path between two points of ${\cal M}$. For sub-Riemannian geometry see \cite {GE, ST,GR,P1,P2,P3,P4}. 

{\it The Heisenberg group, $H^1$}, is a simple simply con\-nec\-ted solvable Lie group  homeomorphic  to the $3$-dimensional Euclidean space, just as the covering group ${\overline {SL(2,{\Re})}}$ of  $SL(2,{\Re})$. The Heisenberg group is the prototype of SR-manifold. It has the structure of a line bundle over the Euclidean plane ${\Re}^2\times {\Re}$; ${\overline {SL(2,{\Re})}}$ is a line bundle over the {\it hyperbolic} Poincar\'e disc  $D_2\times {\Re}$, the prototype of CR-manifold {\it with boundary} ${\cal M}=int \overline {{\cal M}}$. While Aut(${\overline {SL(2,{\Re})}}$) is the symmetry group of the interior of ${\cal M}$, $Aut({\overline G})$ is the symmetry group of the boundary. It is this quintessential principle, connecting the CR-geometry with the SR-geometry, which enable us to identify the SR-manifold of the compact Heisenberg group with the compact boundary of G\"odel's universe. For Heisenberg group see \cite {OW,GB}. For CR-manifolds in general relativity, see Penrose \& Rindler \cite {PEN}. 

The  $
(2n+1)$-dimensional Heisenberg group, $H^{n}$, may be present as the multiplicative group of real
  $(n+2)\times (n+2)$ matrices of the form 
   
\begin{equation}A(x,y,t)=\left( \begin{array}{ccc}
1& x&t\\
0&1& y\\
0&0&1
\end{array}\right)\equiv  (x,y,t)\in {\Re}^{2n}\times {\Re},
\end{equation}
 \noindent The unbounded center is the real line $(0,0,t)\in {\Re}^3$.

 If we identify ${\Re}^{2n}$ with $C^n$, by setting $z=x+iy$ and $w=u+iv$, $H^n$ can be defined by the group law
\begin{equation}
\label{UY}
(z,t)\cdot (w,s)=(z+w,t+s+\frac{1}{2}Im(z\cdot {\overline w})),
\end{equation}
where $Im(z\cdot {\overline w})=(u\cdot y -v\cdot x)$ is the standard symplectic form on ${\Re}^{2n}$ and $z\cdot {\overline w}=z_1{\overline w}_1+\ldots + z_n{\overline w}_n$ is the standard Hermitian form on $C^n$.

The $(2n+1)$-dimensional Heisenberg manifold is given by a contact $1$-form ${\Omega}$ such that ${\Omega}\wedge (d{\Omega})^n \neq 0 $. It is well known the a $(2n+1)$-dimensional manifold given by a contact structure {\it is always orientable if $n$ is {\it odd}}. See Kobayashy \& Nomizu  \cite {KOB}. We shall work with the $n=1$ case.

 The left invariant vector fields, 
\begin{equation}
Q=\frac {{\partial}}{{\partial}_{x}}-\frac {1}{2}y\frac {{\partial}}{{\partial}_{t}}; P=\frac {{\partial}}{{\partial}_{y}}+\frac{1}{2}x\frac {{\partial}}{{\partial}_{t}};\; T=\frac {{\partial}}{{\partial}_{t}}
\end{equation}

\noindent are the generators of the Heisenberg Lie algebra $h^1$:
\begin{equation}
\label{PE}
[P,P]=[Q,Q]=0;[Q,P]=T.
\end{equation}
 
The Heisenberg group $H^1$ can be identified with ${\Re}^3$ by the  exponential map $ exp : h^1\rightarrow H^1\approx {\Re}^3$, 

\begin{equation}
e^{xQ}e^{yP}e^{t[Q,P]}\rightarrow (x,y,t).
\end{equation}

Every point in ${\Re}^3$,(except the origin), can be reached from any other point in ${\Re}^3$, only by   horizontal curves that is: curves whose tangent vectors are $(Q,P)$. Such curves in general are not locally minimizer of the proper time, just like Carnot's adiabatic cycles of classical thermodynamics.

{\it The Heisenberg  motion group} is the semi-direct pro\-du\-ct $ G=H^1 \times U(1)$. This group is the group of isometries of the sub-elliptic operator, (sub-Laplacian ) ${\cal L}$: 

\begin{equation}
\label{BU}
{\cal L}=-(P^2+Q^2)
=-{\Delta}_z-\frac{1}{4}{\bf r}^2{\partial}_t^2+J{\partial}_t,
\end{equation}
\noindent where ${\Delta}_z$ is the Laplacian in the complex plane, ${\bf r}^2=|z|^2$, and 
\begin{equation}
J= (x{\partial}_{y}-y{\partial}_{x})
\end{equation}
the rotation operator.

 In order to have a  boun\-ded  spacetime we must work with  the compact homogeneous group  $H^1/{\Gamma}$. The  discrete compact  sub-group ${\Gamma}$ is generated by the set of three matrices ${\Gamma}=\{(1,0,0),(0,1,0),(0,0,1/2k{\pi});k=\pm 1,\pm 2,\pm 3,...\}$. 

{\it G\"odel's Universe }
  is a rotating anisotropic  homogeneous Lorentzian (hyperbolic) spacetime, in which  matter takes the form of a pressure-free perfect fluid ($T_{ab}={\rho}{\bf u}_a{\bf u}_b)$ where ${\rho}$ is the matter density and ${\bf u}_a$ are the four normalized velocity vector fields. The  important spacetime manifold is ${\cal M}=\Re^3$, with  metric 
\begin{equation}
\label{S}
ds^2=-c^2dt^2 +dx^2 -\frac{1}{2}e^{2{\sqrt 2}{\omega}_0x}dy^2-2e^{{\sqrt 2}{\omega}_0x}cdtdy 
\end{equation}
where ${\omega}_0$ is the angular velocity of the matter field associated with the velocity field ${\bf u}_4$. 

Einstein's field equations are satisfied if $|{\bf u}_{0}|=|{\bf u}_4|=1$ and $4{\pi}G{\rho}={\omega}_0^2=-{\Lambda}$, where $G$ is the gravitational constant. 
 
It is easy to verify, (with a little help of $({\omega}_0,c)$), that G\"odel's apacetime manifold is isometric under local actions of the automorphisms of the Heisenberg motion group:

\vspace{0.15cm}
\noindent 1) Symplectic group $Sp(2,{\Re})$:${\alpha}_1({\bf x})= 
(y,-x,t) \rightarrow ({\omega}_0,-c)$; that is: ${\bf J}(x,y)=(y-x);{\bf J}^2=-1$ 

\noindent 2) Dilation: 
 ${\alpha}_2({\bf x})=(ax,ay,a^2t),\rightarrow ({\omega}_0/a,c/a); a>0$.

\noindent 3)Temporal Inversion: 
${\alpha}_3({\bf x})=(x,y,-t) \rightarrow ({\omega}_0,-c)$ 

\noindent 4) Spatial Inversion:
${\alpha}_4({\bf x})=(-x,-y,t) \rightarrow (-{\omega}_0,-c) $

\noindent 5)  ${\alpha}_5={\bf Z}/2{\bf Z}\oplus{\bf Z}/2{\bf Z}$, consisting of transformations which change sign of any two coordinates.

\noindent 6)Rotation:
 ${\alpha}_6({\bf x})=({\sigma }(x,y),t);\; {\sigma}\in U(1)$ 

\noindent 7)Screw translation:
${\alpha}_7({\bf x})=({\bf R}_{\theta}(x,y),t+ const({\theta}))$,
where ${\bf R}_{\theta}\in SO(2)$ 

 Recall that if two models $({\cal M}_1,d_1)$ and $( {\cal M}_2,d_2)$ have the same local symmetry group, they have similar shape  and they belong to the same equivalence class. As it is well known {\it the local symmetry group dictate the form of the laws of nature}.

So this isometry enable us to identify the orbit space of the compact Heisenberg group with the boundary of G\"odel Universe.

 We shall  work with 
the most symmetric metric representative of the orbit space of the Heisenberg group:
\begin{equation}
\label{A}
ds^2= dx^2 +dy^2 +cdt[cdt-g(xdy-ydx)] 
\end{equation}
\noindent 
\noindent where $g=({\sqrt 2}/2){\omega}_0$.
 
The contact form  ${\Omega}=cdt -g(xdy-ydx)$ is  the annihilator of the vector fields $(Q,P)$, spans of the horizontal sub-space of the tangent space $T{\cal M}=H\oplus V$:
\begin{equation}
 Q={\partial}_x-\frac{gy}{c}{\partial}_t; P={\partial}_y+\frac {gx}{c}{\partial}_t,
\end{equation}
\noindent and the dual of the vector fields, $(T,J)$, spans of the vertical sub-space :
\begin{equation}
T=\frac {1}{c}{\partial}_t ;\;\;J=g(y{\partial}_x-x{\partial}_y).
\end{equation}
\noindent That is,
\begin{equation}
Q({\Omega})=P({\Omega})=0;
T({\Omega})=1; J({\Omega})=g^2(x^2+y^2).
\end{equation}
Einstein's field equations are satisfied if $|T|=|J|=1$. Hence,$|c|=1$ and
\begin{equation}
\label{OI}
g^2(x^2+y^2)=1;\rightarrow {\omega}_0= \frac{{\sqrt 2}}{{\bf r}_0};\rightarrow {\Lambda}=-\frac {2}{{\bf r}_0^2}
\end{equation}

The  integral lines of the vector fields $(Q,P,T,J)$  are solutions of Einstein's field equations. In what follows, we will use the geometrized units ($c=G={\hbar}=1)$. The solutions  depend on the maximal radius of the spacelike surface, ${\bf r}_0^2=1/2{\pi}{\rho}$. The {\it spectrum}  as well as the {\it normal modes} ${\omega}_n={\omega}_n({\omega}_0)$, will be obtained from the {\it spectral radius} ${\bf r}={\bf r}({\bf r}_0$), as we shall see. 

\newpage
{\it In the phase space}, $T^{*}{\cal M}$, the 
 vector fields $(Q,P,T,J)$ can be re\-pre\-sen\-ted  by the functions: 
\begin{equation}
Q=P_x-gyP_t;P=P_y+gxP_t ;T=P_t;J=g(yP_x-xP_y)
\end{equation}

An explicit calculation gives us the Lie algebra:
\begin{equation}
\{Q,P\}=2gT;\{J,P\}=gQ;\{J,Q\}=-gP 
\end{equation}

\begin{equation}
\{P,T\}=\{Q,T\}=\{J,T\}=0.
\end{equation}

This  Lie algebra  arise in the Napi-Witten mo\-del of a four-dimensional homogenous anisotropic spacetime, constructed from an ungauged  Wess-Zumino-Witten model, based on nonsemisimple groups \cite{NA}.

 It will be shown in the following that the metric (\ref {A}) gives us the wave front of a  cylindrical gra\-vi\-ta\-ti\-onal wave, first considered  by Einstein \& Rosen  \cite {AE} and latter des\-cri\-bed in details by Weber \& Wheeler \cite {WE}, and  Marder \cite {MAR}.

The  Cauchy problem has unique solution  from the Hamiltonian
\begin{equation}
 {\cal H}=\frac{1}{2}(Q^2+P^2+T^2)+J^2
\end{equation}
 with the supplementary conditions
\begin{equation}
\label{P}
\frac {dx}{ds}=Q;\;\;\;
\frac {dy}{ds}=P;\;\;\;
\frac {dt}{ds} -g\left(x\frac {dy}{ds}-y\frac {dx}{ds}\right)=0.
\end{equation}

The Hamiltonian equations are
\begin{mathletters}
\begin{equation}
\label{L}
\frac {dQ}{ds}=\{{\cal H},Q\}=P\{P,Q\}+2J\{J,Q\}=-2g{\lambda}P,
\end{equation}
\begin{equation}
\frac {dP}{ds}=\{{\cal H},P\}=Q\{Q,P\}+2J\{J,P\}=2g{\lambda}Q,
\end{equation}
\begin{equation}
\frac{dT}{ds}=\{{\cal H},T\}=0;\;\;\;
\frac{dJ}{ds}=\{{\cal H},J\}=0,
\end{equation}
 \end{mathletters}

Introducing 
$Q={\bf r}_0cos{\psi},P={\bf r}_0sin{\psi}$,
we have 
\begin{equation}
-2g{\lambda}P=\frac {dQ}{ds}=-{\bf r}_0sin{\psi}\frac {d{\psi}}{ds}.
\end{equation}
It follows that ${\psi}= 2g{\lambda}s+{\phi}$, where ${\phi}$ gives the initial directions of the sub-Riemannian light rays emanate from the origin $(0,0,0)$ of the spacetime.
 Integrating (\ref {P}), in the interval $(0,1)$, one obtains
 the sub-Riemannian wave front, that is, the manifold of the endpoints of the sub-Riemannian rays of length ${\bf r}_0$ :

\begin{mathletters}
\begin{eqnarray}
\label{u}
x={\bf r}_0\left[\frac{cos{\phi}-cos({\theta} +{\phi})}{{\theta}}\right],\\
y={\bf r}_0\left[\frac{sin({\theta} +{\phi})-sin{\phi}}{{\theta}}\right],\\
 t={\bf r}_{0}^2\left[\frac{{\theta} -sin {{\theta}}}{{\theta}^2}\right],
\end{eqnarray}
\end{mathletters}
\noindent with ${\theta}= 2g{\lambda}={\sqrt 2}{\omega}_0{\lambda}$, where, ${\lambda}\in {\Re}$, is the {\it spectral parameter} and  $0\leq {\phi}\leq 2{\pi}$. 
These spherical harmonic functions belong to the commutative Banach subalgebra $L^{1}({\overline G}/U(1))$ of the noncommutative algebra $L^1({\overline G})$ of the compact Heisenberg motion group. The pair $({\overline G},U(1))$ is known as Gelfand pair in harmonic analysis on the Heisenberg group see \cite {SUT}.It is easy to verify, aided by computers, that these parametric equations have just the $7$-symmetries  of  $Aut({\overline G}/U(1))$. These equations gives
  all relevant information about the {\it geometry, topology and spectral representation} of the holographic boundary of $(2+1)$-dimensional G\"odel spacetime manifold:
\begin{enumerate}
\item The axial symmetry,the anisotropic dilation and mirror symmetry give us the  spectral distance function from the origin $(0,0,0)$ as $d({\bf 0},(x,y,t))= |x|+|y|+ \sqrt {|t|}$ that is:
\begin{equation}
\frac {d({\bf 0},(x,y,t))}{a{\bf r}_0}=f({\theta})= \left( \frac {(1-cos{\theta})}{{\theta}} + \frac {sin{\theta}}{{\theta}}+ \frac {\sqrt {{\theta}-sin{\theta}}}{{\theta}}\right)
\end{equation}
This function is symmetric $f({\theta})=f(-{\theta})$, positive definite and finite, such that $f(0)=1$ and $f({\theta})\rightarrow 0$ as ${\theta}\rightarrow {\infty}$, with local maxima at ${\theta}=({2n+1}){\pi}$, ($n=0,1,2...$), as  shown in fig.1.
\item The spectral radius is given by ${\bf r}=\sqrt {x^2+y^2}$, or 
\begin{equation}
{\bf r}={\bf r}_0\frac {\sqrt {2(1-cos{\theta})}}{{\theta}} 
\end{equation}
\item The normal modes are given by the local maxima of the spectral radii ${\bf r}_{2n+1}$, at ${\theta}=(2n+1){\pi}$:
\begin{equation}
{\omega}_{2n+1}= \frac{{\sqrt 2}}{{\bf r}_{2n+1}}={\omega}_0(n+ \frac {1}{2}){\pi};\;\;\;n=0,1,2,...
\end{equation}
\item The future and past horizons, $H^{\pm}$, located in the regions $\pm {\pi}\leq {\theta}\leq  \pm 2{\pi}$, has maximal radii at the edge of the spacetime, $({\bf r}_{\pm}=2{\bf r}_0/{\pi}, t_h=\pm {\bf r}_0^2/{\pi})$.
\item The refocusing incomplete geodesic formed by an infinite set of rectifiable curves, is shown in fig.2.
\begin{figure}
\begin{center}
\includegraphics[width=0.3\textwidth]{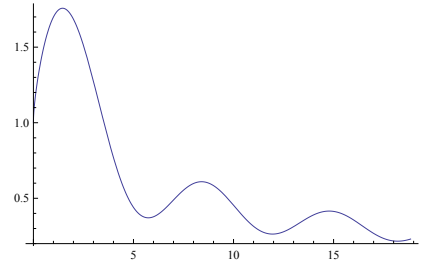}
\caption{ The anisotropic spectral distance from the origin in G\"odel's spacetime, from ${\theta}=0$ to ${\theta}=6{\pi}$} 
\end{center}
\end{figure}
\begin{figure}
\begin{center}
\includegraphics[width=0.2\textwidth]{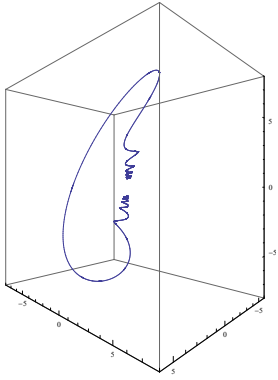}
\caption{ The incomplete inextendible geodesic from ${\theta}=-10{\pi}$ to ${\theta}=10{\pi}$} 
\end{center}
\end{figure}

\begin{figure}
\begin{center}
\includegraphics[width=0.2\textwidth]{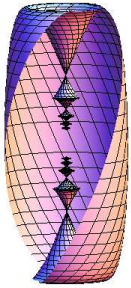}
\caption{The Cauchy spacelike surface (with horizons),  from ${\theta}=-10{\pi}$ to ${\theta}=10 {\pi}$} 
\end{center}
\end{figure}
\begin{figure}[!h]
\begin{center}
\includegraphics[width=0.2\textwidth]{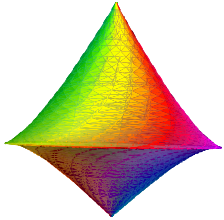}
\caption{ The first G\"odel universe (without horizons) in the region $2{\pi}\leq {\theta} \leq 4{\pi}$.(To be compared with fig.31 in \cite {HW})}\_. 
\end{center}
\end{figure}
\end{enumerate}

{\it Conclusions} - In this  letter we have presented a global cylindrical solution for the geodesic motion in the $(2+1)$ - dimensional  G\"odel spacetime, based in the geometry of the compact Heisenberg motion group. Cau\-sa\-li\-ty violation is not vi\-si\-ble from outside due to the existence of a well  defined closed Cauchy-Riemann spacelike surface, ${\Sigma}={\Sigma}^{+}\cup {\Sigma}^{-}$, which hosts two mirror sym\-me\-tric  sets of countable self-similar non causal (not globally hyperbolic) G\"odel's surfaces. The global spacetime is {\it glo\-bally hyperbolic} and causally geodesically incomplete (according to Hawking-Penrose conditions \cite {JS}) and strongly supports Hawking's chronology protection conjecture:{\it the laws of physics do not allow the appearance of closed timelike curves}.

There is no negative energy states. As it is well know, in $3$-dimensional Einstein gravity, the only possible e\-ner\-gy  measure must be topological and the Euler invariant $e({\Sigma})$ is the only candidate, according to the three-dimensional gravity theory of Deser, Jackiw and 't Hooft \cite {DE}. As is well known, after Thurston \cite {TU}, in the case of a closed $3$-manifold with  geometric structure modeled on the compact Heisenberg group $H^1/{\Gamma}$, $e({\Sigma})=1$. 

 The internal geometry is fixed by the set of focal points of the long geodesic, which is not spacelike everywhere, as shown in fig.2. The source's world line is the {\it locci} of focal points of the internal timelike geodesic. This closed homogeneous ani\-so\-tro\-pic spa\-ce\-time shares many in\-te\-res\-ting properties with a physically acceptable spacetime. However, due to the notable absence of the affine con\-nec\-tion, G\"odel's spacetime  is not an ac\-ce\-pta\-ble spacetime in the con\-ven\-ti\-onal ge\-ne\-ral theory of relativity exposed in Misner,Thorne \& Wheeler \cite {MIS} and in Weinberg \cite {WEI}. Indeed, the me\-tric structure of the rotating body, (fig.3), is rather complicated such that it cannot be isometrically imbedded in the Euclidean spa\-ce ${\Re}^3$. One can say  that the set of end points of the sub-Riemannian rays emanate from the origin is a convex set of points in, ${\Re}^3$, whose external  boundary has  prolongation to the non-trivial internal spacetime, which has no end point. The external boundary and the internal spa\-ce\-time are mediated by two  censor horizons at the edge of the spacetime. The only really singular point is the un\-re\-a\-cha\-ble origin $(0,0,0)$  {\it fons et origo } and  {\it closure}  of the whole spacetime, {\it not visible from outside.} 

 There is a close connection between this approach and \cite {DD}, where the flat Kerr metric was used to represent the spacetime. Indeed, the three-dimensional manifold of the compact Heisenberg group, $H^1/{\Gamma}\approx {\Re}^2\times {\Re}/{\Gamma}$, is a line  bundle over a plane, in ${\Re}^3$, with a prescription for identifying points in ${\Re}^3$. This $3$-dimensional topology was used in, \cite {DD}, to give an example of how to enclose the spacetime in a closed  Cauchy-Riemann surface:

{\it "The matching condition is defined only when a closed curve is followed around the source ;these matchings are defined  by a deficit angle and a time shift. More precisely we have a space with ${\Re}^3/{\Re}$ topology. (a three-space  with a line obstruction on the source's world line) and a prescription for identifying points."} Op.Cit
\cite {DD}.

These matchings are defined by the automorphisms of the compact Heisenberg group ${\overline G}/U(1)=H^1/{\Gamma}$.

It should be noted that, the topology of the external boundary is trivial in the region $-2{\pi}\leq {\theta}\leq 2{\pi}$, that is ${\pi}_1(S^2)=0$. However, it is not contractible, because the internal spacetime, beyond the horizons, is non trivial since  ${\pi}_2({\overline G}/U(1))={\pi}_1(U(1))={\bf Z}$

So, there exist a kind of second order (discontinuous) Landau-Guinsbourg  phase transition from  trivial to nontrivial topology mediated by a first order (continuous) phase transition at the edge of the horizons.

The global (non-pertubative) quantum theory can be obtained by means the celebrated  Stone-Von Neumann theorem on the unitary and irreducible representation of the Heisenberg group in the Hilbert space, which is fairly simple and well understood \cite {OW,GB,SUT}.

These  results provide an appropriate topological  sce\-na\-rio for non-Abelian gauge theories, as : Gribov's theory of quark confinement \cite {GRI}, massive $3$-D gravity theory \cite {JAW} and $3$-D to\-po\-lo\-gi\-cal  insulators,\cite {TA,KA,DC}.

\section*{\bf Acknowledgments}I am grateful to S.Deser, S.Jofilly, L.Ryff, K.Mundim, A.Antunes, C.Sigaud, C.Farina and J.Koiller for discussions, and to Marcela Gon\c calves for encouragements.

\end{document}